\documentclass[12pt,preprint]{aastex}

\lefthead{Chou et al.}
\righthead{Chemically Testing the Origin of
GASS}

\begin{document}
\title{The Chemical Evolution \\
of the Monoceros Ring/Galactic Anticenter Stellar Structure}

\author{Mei-Yin Chou\altaffilmark{1,2},
Steven R. Majewski\altaffilmark{2}, Katia Cunha\altaffilmark{3,4},
Verne V. Smith\altaffilmark{3}, \\
Richard J. Patterson\altaffilmark{2}, and David
Mart{\'i}nez-Delgado\altaffilmark{5,6}
 }

\altaffiltext{1}{Institute of Astronomy and Astrophysics, Academia
Sinica, Taipei 10617, Taiwan (cmy@asiaa.sinica.edu.tw)}

\altaffiltext{2}{Dept. of Astronomy, University of Virginia,
Charlottesville, VA 22904-4325 (mc6ss, srm4n, rjp0i@virginia.edu)}

\altaffiltext{3}{National Optical Astronomy Observatories, PO Box
26732, Tucson, AZ 85726 (cunha, vsmith@noao.edu)}

\altaffiltext{4}{On leave from Observatorio Nacional, Rio de
Janeiro, Brazil}

\altaffiltext{5}{Instituto de Astrofisica de Canarias, La Laguna,
Spain (ddelgado@iac.es)}

\altaffiltext{6}{Max-Planck-Institut f\"{u}r Astronomie,
K\"{o}nigstuhl 17 , Heidelberg, Germany }

\begin{abstract}

The origin of the Galactic Anticenter Stellar Structure (GASS) or
``Monoceros ring'' --- a low latitude overdensity at the edge of
the Galactic disk spanning at least the second and third Galactic
quadrants --- remains controversial. Models for the origin of GASS
fall generally into scenarios where it is either a part (e.g.,
warp) of the Galactic disk or it represents tidal debris from the
disruption of a Milky Way (MW) satellite galaxy. To further
constrain models for the origin of GASS, we derive chemical
abundance patterns from high resolution spectra for 21 M giants
spatially and kinematically identified with it. The abundances of
the (mostly) $\alpha$ element titanium and s-process elements
yttrium and lanthanum for these GASS stars are found to be lower
at the same [Fe/H] than those for MW stars, but similar to those
of stars in the Sagittarius stream, other dwarf spheroidal
galaxies, and the Large Magellanic Cloud. This demonstrates that
GASS stars have a chemical enrichment history typical of dwarf
galaxies --- and unlike those of typical MW stars (at least MW
stars near the Sun). Nevertheless, these abundance results cannot
definitively rule out the possibility that GASS was dynamically
created out of a {\it previously formed}, outer MW disk because
$\Lambda$CDM-based structure formation models show that galactic
disks grow outward by accretion of dwarf galaxies. On the other
hand, the chemical patterns seen in GASS stars do provide striking
verification that accretion of dwarf galaxies has indeed happened
at the edge of the MW disk.

\end{abstract}

\keywords{Galaxy: evolution --- Galaxy: halo --- Galaxy: structure
--- galaxies: dwarf --- galaxies: interactions --- stars: abundances }

\section{Introduction}

The Galactic Anticenter Stellar Structure (GASS), also called the
``Monoceros Stream'' or ``Monoceros Ring'', was discovered as
several overdensities of presumed main sequence turnoff stars with
F-type colors in the Sloan Digital Sky Survey (SDSS; Newberg et
al. 2002) at a mean distance of $\sim$$11$ kpc from the Sun in the
general region of the Galactic anticenter direction in Monoceros.
Subsequent study of this arc-like structure with 2MASS M giant
tracing (Majewski et al. 2003, hereafter M03; Rocha-Pinto et al.
2003, hereafter RP03) and spectroscopy (Crane et al. 2003,
hereafter C03), with the Isaac Newton Telescope Wide Field Camera
(INT/WFC) (e.g., Ibata at al. 2003, hereafter I03), and with SDSS
photometry and spectroscopy (e.g., Yanny et al. 2003, hereafter
Y03) has shown the low latitude GASS ring spans at least the
second and third Galactic quadrants and a wide metallicity range,
from [Fe/H]= $-1.6\pm 0.3$ (Y03) to $-0.4\pm 0.3$ (C03).

The origin of GASS remains controversial. Models for its formation
fall into one of two general categories: (1) scenarios where GASS
is a part of the Milky Way (MW) disk seen at higher latitudes due
to the disk warp or some other dynamical excitation of the disk
(Ibata at al. 2003; Momany et al.2006; Kazantzidis et al. 2008;
Younger et al. 2008), versus (2) those scenarios where GASS
represents a tidal debris stream from the disruption of a MW
satellite galaxy (Ibata at al. 2003; Y03; C03; RP03; Helmi et al.
2003; Frinchaboy et al. 2004; Martin et al. 2004a; Conn et al.
2007; Pe{\~{n}}arrubia et al. 2005; Rocha-Pinto et al. 2006).
Furthermore, Martin et al. (2004a, 2004b) claim to find an
overdensity in Canis Major (CMa, $l \sim 240^{\circ}$) and assume
it is the progenitor of GASS, although the existence of this
overdensity itself has been challenged (e.g., Momany et al. 2006;
Rocha-Pinto et al. 2006; Mateu et al. 2009). While several
candidates for a parent system to the putative tidal stream have
been proffered (Martin et al. 2004a, 2004b; Rocha-Pinto et al.
2006), the basic issue of a disk versus stream origin is still
hotly debated (Martin et al. 2004b; Momany et al. 2006; Conn et
al. 2007). It is interesting to note that Bonifacio et al. (2000)
suggested that a small galaxy might be hidden in this region of
the Galaxy based on their derived distance to Nova V 1493 Aql.

In this Letter we gain insights into the origin of the GASS system
by exploring its chemistry. Because chemical enrichment histories
are a function of environment as well as other, stochastic
processes, it is expected that stars that formed originally in
different, isolated environments will bear significantly different
chemical imprints in their spectra. Such differences have already
been demonstrated by, for example, depressed $\alpha$-element and
s-process abundances in dwarf galaxies compared to nearby MW stars
(e.g., Shetrone et al.\ 2003; Venn et al.\ 2004; Geisler et al.\
2005; Chou et al. 2010, hereafter C10). Here we analyze the
abundances of the primarily $\alpha$-element titanium (Ti) and
s-process elements yttrium (Y) and lanthanum (La) in GASS M giant
candidates, and we compare those abundances with those of the
Sagittarius (Sgr) dwarf galaxy, the Large Magellanic Cloud (LMC)
and other dwarf spheroidal (dSph) satellite galaxies of the MW
(including Sculptor, Carina, Draco, Ursa Minor, Fornax, Sextans
and Leo I). A comparison to the Sgr and LMC systems is
particularly apropos in the study of GASS because Sgr and the LMC
exhibit some similar properties to GASS, including relatively high
metallicities and the presence of M giants (C03; M03; RP03; C10).
It is interesting to know whether these similarities between the
GASS, Sgr and LMC systems extend to their detailed chemical
patterns, or if GASS chemistry more closely resembles that of the
typically more metal-rich MW disk.

\section{Observations and Analysis}

We obtained high resolution spectra of 21 stars spatially
coincident (i.e. by position on the sky and projected distance)
with the GASS system based on M03's work, but, additionally,
vetted for proper GASS radial velocity by C03. The data were taken
with the Echelle spectrograph mounted on the Kitt Peak National
Observatory Mayall 4-m telescope and the SARG spectrograph on the
Telescopio Nazionale Galileo (TNG) 3.5-m telescope\footnote{The
Italian Telescopio Nazionale Galileo is operated on the island of
La Palma by the Fundacion Galileo Galilei of the INAF (Istituto
Nazionale di Astrofisica) at the Spanish Observatorio del Roque de
los Muchachos of the Instituto de Astrofisica de Canarias.},
having resolutions of $R=35,000$ and 46,000, respectively.
Examples of spectra from each instrument can be seen in Figure 3
of Chou et al. (2007, hereafter C07) and Figure 1 of C10.

Analysis of the spectra follows very closely that used in the
analysis of Sgr system M giants described in C07 and C10. We focus
on measuring equivalent widths (EWs) of eleven \ion{Fe}{+1} lines,
two \ion{Ti}{+1} lines, and one \ion{Y}{+2} line, and analyzing by
way of spectral synthesis one \ion{La}{+2} line; these lines all
lie in a particular part of the spectrum (7440-7590 \AA\ )
previously investigated by, e.g., Smith \& Lambert (1985) in their
spectroscopic exploration of M giants.

Our derivation of the abundances of these elements uses the LTE
code MOOG (Sneden 1973) and follows the same procedures described
in C07 and C10, where the excitation potentials and $gf$-values
adopted for each line are also given. These particular chemical
elements were chosen not only because they have well-defined,
measurable spectral lines in M giants, but also because they show
distinctive abundance ratios (relative to Fe) between different
dwarf galaxies and, as a group, when these dwarf systems are
compared to the MW.

The model atmospheres adopted here are the same as in C07 and C10,
and were generated by linear interpolation from the grids of
ATLAS9 models (Castelli \& Kurucz 2003).\footnote{From
http://kurucz.harvard.edu/grids.html.} We used the ODFNEW models
with a microturbulence of 2 km s$^{-1}$ and no convective
overshooting in our analysis.

Table 1 summarizes the targets, their equatorial and Galactic
coordinate positions, the velocity in the Galactic Standard of
Rest frame ($v_{GSR}$)\footnote{$v_{GSR}=v_{LSR}+220cos(b)sin
(l)=v_{helio}+9cos(b)cos(l)+231cos(b)sin(l)+6sin(b)$, where
$v_{helio}$ is the measured heliocentric radial velocity,
$v_{LSR}$ is the radial velocity with respect to the Local
Standard of Rest, $b$ is the Galactic latitude and $l$ is the
Galactic longitude.}, the spectrograph with which each target was
observed and on what date, and the $S/N$ of each spectrum. The
$v_{GSR}$ values were derived by cross-correlating the echelle
order that we used for the chemical analyses against the same
order for several radial velocity standard stars taken from the
Astronomical Almanac. The estimated standard deviations of the
velocities are $\sim1.2$ km s$^{-1}$ for the KPNO spectra, and 0.5
km s$^{-1}$ for the SARG spectra. The $S/N$ was determined using
the total photoelectron count level at 7490\AA. Table 2 lists the
apparent magnitudes $K_s$, dereddened $(J-K_s)_o$ colors, and the
derived abundance results for these GASS stars. The columns in
Table 2 give the derived effective temperature using the
Houdashelt et al. (2000) color-temperature relation applied to the
2MASS $(J-K_s)_o$ color, and the derived values of the surface
gravity ($\log{g}$), microturbulence ($\xi$), abundance $A$(X),
and abundance ratios [Fe/H] or [X/H] for each element X as well as
the standard deviation in the abundance determinations. The
microturbulence is determined by minimizing the dependence between
the derived Fe abundance and the Fe I EW for the different lines.
The standard deviation represents the line-to-line scatter of the
EW measures (for Fe, Ti and Y), or different continuum level
adjustments (for La), as discussed in C07 and C10. The final
measured EWs of the lines we analyzed for each of the GASS spectra
will be given elsewhere (M.-Y. Chou et al., in prep.).

Abundance uncertainties were estimated for three sample GASS stars
by varying the model atmosphere in effective temperature ($T_{\rm
eff}$), surface gravity (log $g$) and microturbulent velocity
($\xi$). The sensitivity of abundances to these stellar parameters
is given in Table 3. The uncertainty in effective temperature is
$\sim$$\pm80$ K based on the Houdashelt et al. (2000a,b)
color-temperature relation. The uncertainty in the surface gravity
is $\sim$$\pm0.15$ dex, which also comes from the uncertainty in
the effective temperature, propagated along the corresponding
isochrone (see C07). The uncertainty in microturbulent velocity is
$\sim$$\pm0.1$ km s$^{-1}$, which we estimate from the relation of
abundance $A$(Fe) versus Fe I EW for different lines. The final,
estimated errors propagated from these uncertainties in the
stellar parameters are 0.08, 0.11, 0.07, 0.05 for [Fe/H], [Ti/Fe],
[Y/Fe], and [La/Fe], respectively. Finally, the EW measurement
uncertainty is $\sim3$-$5$ m\AA\ , estimated by the equation given
in Smith et al. (2000). Combining the uncertainties in stellar
parameters and EW measurements, we estimate the net abundance
errors to be no more than $\sim0.2$ dex for each element ratio in
Table 2.

\section{Results and Discussion}

Figure 1 compares the derived distributions of [Ti/Fe], [Y/Fe],
and [La/Fe] as a function of [Fe/H] for the GASS, Sgr stars (the
latter from C07 and C10), those of the LMC and other dSph
galaxies. The lines represent linear fits to the abundance
distributions of nearby disk and halo stars (see C10). As may be
seen in the left panels, in all cases (1) the nearby MW and GASS
abundance patterns are not coincident, while (2) the distribution
of abundances and metallicities for GASS and Sgr stars is very
similar, except for the [La/Fe] patterns at the highest
metallicities (where the relative patterns suggest that Sgr stars
probably enriched somewhat more slowly than GASS stars). The
overall similarity of GASS and Sgr patterns shows that the stars
in these two systems had a rather similar enrichment history, and
may provide at least circumstantial evidence that the GASS/Mon
stars originally derived from an accreted dwarf galaxy similar to
that represented by the Sgr dwarf galaxy.

On the other hand, Sbordone et al. (2005) showed the chemical
abundances for three giant star candidates of the CMa overdensity.
The higher [Ti/Fe] and [La/Fe] of these CMa stars compared to
those from GASS in the left panels of Figure 1 may suggest the CMa
overdensity is not part of the GASS. The chemistry of these CMa
stars more closely resembles that for the Sgr dSph, but,
interestingly, also differs from that of Galactic disk stars (see
the comparison with the Galactic trends). Though more data on CMa
stars are certainly warranted, the results shown here support the
conclusion that CMa and GASS are distinct from one another (e.g.,
Mateu et al. 2009).

However, we note that these CMa stars for which Sbordone et al.
(2005) have measured abundances are K giants.\footnote{The same is
true for some of the Sgr stars in Figure 2 that we have included
from Sbordone et al. (2007).}. Some of the systematic differences
between the Ti abundances derived for K and M giants in the
different studies could partially come from the fact that these
abundances were computed in LTE. In addition to possible
differences due to NLTE effects in \ion{Ti}{+1} transitions
between K and M giants, it is worth noting that the \ion{Ti}{+1}
lines used by Sbordone et al. (2005) and this study are all
different, making detailed comparisons subject to some uncertainty
due to using different spectral lines, which may have different
NLTE corrections.

In the right panels of Figure 1 it is obvious that GASS goes to
higher [Fe/H] than the LMC and especially the other dSphs, but
most stars in all of these systems have similarly low [Ti/Fe] and
[Y/Fe] compared to the MW. This resemblance is reinforced by
Figure 2, which shows the distribution of [Y/Fe] versus [Ti/Fe]
for GASS, Sgr, other dSph, LMC and MW stars. As may be seen, the
distribution of most of the GASS stars in this observational plane
resembles those for Sgr, other dSphs and the LMC, which indicates
they share the same chemical patterns and may have similar
chemical evolution histories. Figure 2 also further demonstrates
the differences in chemical evolution between the GASS stars and
the nominal MW populations.

On the other hand, the [La/Fe] ratios for most dSph stars, as well
as more metal-rich LMC and Sgr stars, are enhanced compared to the
MW, in strong contrast to what is observed among our (metal-rich)
GASS stars. In this way GASS distinguishes itself from all other
MW satellites with known [La/Fe] measures.

As discussed in C10, the distinctive enrichment history of a
stellar system leads to particular abundance patterns in its
stars, and therefore ``chemical fingerprinting'' can help identify
tidally stripped and captured stars in the Galactic field. For
instance, $\alpha$-elements are mainly produced from Type II
supernovae (SN II) while iron is synthesized largely by Type Ia
supernovae (SN Ia). So [$\alpha$/Fe] is high in the early
enrichment process of a stellar system until SN Ia ignite after
the first $\sim$$1$ Gyr. Ti acts primarily as an $\alpha$ element
(although it also can be produced in SN Ia). The chemical patterns
of the $\alpha$-elements (e.g., Ti) and s-process elements (e.g.,
Y and La) may demonstrate differences in the (early) star
formation rate (SFR) between the progenitor GASS, Sgr, other dSph,
LMC and MW systems.

The straight arrows on Figure 2 illustrate schematically which way
these various supernovae yields move the combinations of element
ratios in that abundance plane. As the straight arrows in Figure 2
show, SN II produce little Y but some Fe, so that [Y/Fe] decreases
in the early enrichment stage, while Ti is made in proportion to
Fe (at a super-solar level). After SN Ia occur, Fe is produced
more prodigiously compared to Ti and other $\alpha$ elements, and
stars initially move to the left and slightly down in Figure 2.
Following the subsequent contribution of low mass AGB yields, the
[Y/Fe] starts to increase with increasing [Fe/H] (see curved
[Fe/H] arrow). The lower [$\alpha$/Fe] trend for GASS compared to
the MW at the same [Fe/H] indicates a slower initial GASS SFR,
which allows much of the iron from SN Ia to be introduced at lower
overall metallicities.  This is a feature that can be seen in many
dSph galaxies as well as the LMC (Figs.\ 1 and 2; also see, e.g.,
Shetrone et al.\ 2003; Venn et al.\ 2004; Geisler et al.\ 2005;
Pomp{\'e}ia et al.\ 2008; C10).

Meanwhile the s-process elements are primarily generated in low
mass asymptotic giant branch (AGB) stars. The delay in the
introduction of significant s-process element yields leads to the
eventual upturn in the evolutionary track (followed by the mean
[Fe/H] level) seen in Figure 2.\footnote{It is worth noting that
the MW stars show the same general evolutionary distribution in
Figure 2, only shifted in mean abundance levels from the satellite
galaxies and GASS.} These AGB stars produce heavier s-process
elements like La more efficiently than lighter species such as Y
in more metal-poor environments (see the review by Busso, Gallino,
\& Wasserburg 1999). C10 has shown underabundant trends in [Y/Fe]
for most Sgr stars compared to the MW, and an upturn in [La/Fe]
for Sgr at [Fe/H]$>$$-0.5$ (Fig.\ 1). The latter trend suggests a
slow enrichment history so that the yields from low-metallicity
AGB stars have enough time to contaminate the interstellar medium
and leave their signature on the metal-rich Sgr populations (Venn
et al.\ 2004; Pomp{\'e}ia et al.\ 2008). The left middle panel in
Figure 1 shows the similarly lower trends of [Y/Fe] for Sgr and
GASS compared with the MW, which may suggest the GASS and Sgr
progenitors are from similar metal-poor environments. However, the
left bottom panel of Figure 1 shows no apparent upturn in [La/Fe]
for GASS stars as seen in Sgr. This indicates that GASS may have
enriched somewhat faster than Sgr, which could explain why we do
not see the low-metallicity AGB yields in the present GASS stellar
population.

In the end, that the chemical characteristics of $\alpha$ and
s-process elements for our GASS stars are very similar to those
seen in Sgr and other satellite galaxies (see Figs.\ 1-2) suggests
a dSph-like environment for the origin of GASS stars, with the
closest matches to the LMC and Sgr as prototypes when overall
metallicity is also considered.\footnote{The inconsistency of
radial velocities and positions between stars in the GASS/Mon
system (C03) and those in either Sgr disruption models (e.g., Law,
Johnston, \& Majewski 2005), observed in the Sgr system (e.g.,
Majewski et al. 2003, 2004), or corresponding to the LMC precludes
an actual physical connection between GASS and these two other
M-giant rich systems, however.} In addition, Lanfranchi, Matteucci
\& Cescutti (2008) predict subsolar [Y/Fe] and [La/Fe] for dSphs
in their standard model, which further reinforces that GASS stars
enriched in a dSph-like environment.

However, we hasten to add that this chemical connection of GASS
stars to a dwarf galaxy origin alone cannot resolve the debate
over whether GASS/Mon is presently a distinct tidal stream or a
warp or other structure generated from a previously formed, outer
MW disk because it is now believed that galactic disks grow
outward by the accretion of dwarf galaxies. For example, the
chemical abundance studies of MW disk red giants and Cepheids by
Yong et al. (2005, 2006) and Carney et al. (2005) suggest that the
outer Galactic disk may have formed via merger events. Numerical
simulations of galaxy evolution in a Cold Dark Matter context also
support that galactic disks grow outward by the accretion of dwarf
galaxies (Abadi et al. 2003; Brook et al. 2007; Read et al. 2008).
Finally, recent studies propose that flyby satellite encounters
could cause ringlike features in the outer disk (Kazantzidis et
al. 2008; Younger et al. 2008). Thus, it may well be that the
outer disk could be made from recently acquired debris of multiple
Sgr-like systems that then through dynamical processes transformed
into what we see as the GASS/Mon structure today. A more
interesting chemical discriminant between the competing GASS/Mon
formation scenarios outlined in \S1 would be to test GASS
chemistry against that of {\it bona fide} outer disk stars (see
M.-Y. Chou et al., in prep.) to see if GASS is pre- or post-outer
MW disk material.

\acknowledgements We appreciate helpful conversations with Ana
Garcia-Perez. M.-Y.C., S.R.M. and R.J.P. acknowledge support from
NSF grants AST-0307851 and AST-0807945. This project was also
supported by the NASA {\it SIM Lite} key project {\it Taking
Measure of the Milky Way} under NASA/JPL contract 1228235. K.C.
and V.V.S. also thank support from the NSF via grant AST-0646790.
D. M.-D. acknowledges funding from the Spanish Ministry of
Education and Science (Ram{\'o}n y Cajal program contract and
research project AYA 2007-65090).

\clearpage
\begin{deluxetable}{lrrrrrccr}
\tabletypesize{\tiny} \tablecaption{The Program Stars}
\tablewidth{0pt} \tablehead{ \colhead{Star No.} &
\colhead{$\alpha$(2000)} & \colhead{$\delta$(2000)} &
\colhead{$l$} & \colhead{$b$} & \colhead{$v_{\rm gsr}$} &
\colhead{Spectrograph} & \colhead{Observation} &\colhead{S/N}
\\
\colhead{} & \colhead{(deg)} & \colhead{(deg)} & \colhead{(deg)} &
\colhead{(deg)} & \colhead{(km s$^{-1}$)} & \colhead{} &
\colhead{UT Date} &\colhead{}
 }

\startdata

GASS &  &  &  &  &  &  &  &\\
$J02340206+8446368$ & 38.50857 & 84.77688 & 125.37 & 22.40 & 56.7 & ECHLR & 2006 Dec 06 & 82\\
$J05144786+8605371$ & 78.69943 & 86.09364 & 126.88 & 25.47 & 66.7 & ECHLR & 2006 Dec 05 & 87\\
$J06044611+3942523$ & 91.19213 & 39.71452 & 172.60 & 8.80 & -13.1 & SARG & 2004 Mar 11 & 40 \\
$J06341077+2421564$ & 98.54485 & 24.36566 & 189.25 & 7.30 & -26.0 & SARG & 2004 Mar 10 & 39\\
$J06402752+5944247$ & 100.11467 & 59.74018 & 155.78 & 21.86 & 20.5 & ECHLR & 2006 Dec 08 & 93\\
$J06463306+5521444$ & 101.63773 & 55.36232 & 160.54 & 21.37 & 57.5 & ECHLR & 2006 Dec 07 & 104\\
$J07003934+4730561$ & 105.16393 & 47.51558 & 169.27 & 21.13 & 27.3 & ECHLR & 2004 May 09 & 47\\
$J07022522+2823272$ & 105.60508 & 28.39090 & 188.24 & 14.73 & -29.7 & ECHLR & 2004 May 07 & 60\\
$J07041061+3911224$ & 106.04421 & 39.18957 & 177.89 & 19.05 & -16.6 & ECHLR & 2004 May 06 & 66\\
$J07052136+6012353$ & 106.33901 & 60.20981 & 156.10 & 25.00 & 7.7 & ECHLR & 2006 Dec 08 & 100\\
$J07074384+1319528$ & 106.93267 & 13.33134 & 202.79 & 9.59 & -30.7 & SARG & 2004 Mar 12 & 38\\
$J07110708+5027276$ & 107.77949 & 50.45766 & 166.73 & 23.59 & 36.8 & ECHLR & 2006 Dec 06 & 112\\
$J07165539+5409046$ & 109.23078 & 54.15129 & 163.01 & 25.31 & 15.3 & ECHLR & 2006 Dec 08 & 95\\
$J07314622+5936181$ & 112.94259 & 59.60503 & 157.34 & 28.17 & 28.9 & ECHLR & 2004 May 08 & 63\\
$J07393712+3315325$ & 114.90466 & 33.25903 & 186.39 & 23.91 & -38.7 & ECHLR & 2004 May 08 & 66\\
$J08205065+1523392$ & 125.21104 & 15.39422 & 208.50 & 26.57 & -44.7 & ECHLR & 2004 May 06 & 82\\
$J08501690+2545291$ & 132.57042 & 25.75807 & 199.75 & 36.54 & -12.1 & ECHLR & 2004 May 07 & 60\\
$J08534628+0604280$ & 133.44281 & 6.07444 & 221.99 & 29.90 & -53.8 & ECHLR & 2004 May 05 & 61\\
$J09122898+2632290$ & 138.12073 & 26.54140 & 200.41 & 41.56 & -21.6 & ECHLR & 2004 May 05 & 63\\
$J09164929+5630048$ & 139.20537 & 56.50134 & 159.79 & 42.17 & 12.0 & ECHLR & 2004 May 05 & 64\\
$J22260469+8218595$ & 336.51956 & 82.31651 & 118.07 & 20.86 & 58.3 & ECHLR & 2006 Dec 06 & 84\\

\enddata
\end{deluxetable}

\begin{deluxetable}{lr@{}rr@{}r@{}l@{}rrrrr@{}lrr@{}lr@{}lr}
\tabletypesize{\tiny} \tablecaption{Stellar Parameters and
Chemical Abundances for the Program Stars} \tablewidth{0pt}
\tablehead{ \colhead{Star No.} & \colhead{$\!K_{s,o}\,$} &
\colhead{$\,(J-K_s)_o\!$} & \colhead{$T_{\rm
eff}$\tablenotemark{a}} & \colhead{log{\it g}\tablenotemark{b}} &
\colhead{} & \colhead{$\xi$ } & \colhead{A(Fe)} &
\colhead{[Fe/H]\tablenotemark{c}} & \colhead{A(Ti)} &
\colhead{[Ti/Fe]\tablenotemark{c}}& \colhead{}& \colhead{A(Y)} &
\colhead{[Y/Fe]\tablenotemark{c}} & \colhead{} & \colhead{A(La)} &
\colhead{} & \colhead{[La/Fe]\tablenotemark{c}}
\\
\colhead{} & \colhead{} & \colhead{} & \colhead{(K)} &
\colhead{(${\rm cm\, s^{-2}}$)} & \colhead{} & \colhead{(${\rm
km\,s^{-1}}$)} & \colhead{} & \colhead{} & \colhead{} & \colhead{}
& \colhead{} & \colhead{} & \colhead{}& \colhead{}& \colhead{} &
\colhead{} & \colhead{}
 }

\startdata
Sun & \nodata & \nodata & \nodata & \nodata && \nodata & 7.45 & \nodata  & 4.90 & \nodata & &  2.21 &  \nodata & & 1.13 & & \nodata   \\
\\
GASS &  &  &&  &  &  &  &  &  &  &  &  &  &  & & \\
$J02340206+8446368$ &  9.83 & 1.07 & 3750 & 0.4&    & 1.37 & 6.87 & $-0.58/  0.12$ & 4.44 & $ 0.12 / 0.06$    && 1.34 & $-0.29 / 0.07$    &&  0.32&\tablenotemark{d} & $-0.34 / 0.02$ \\
$J05144786+8605371$ &  9.57 & 1.05 & 3750 & 0.3&    & 1.43 & 6.73 & $-0.72/  0.10$ & 4.01 & $-0.17 / 0.07$    && 1.34 & $-0.15 / 0.01$    &&  0.23&\tablenotemark{d} & $-0.28 / 0.04$ \\
$J06044611+3942523$ &  9.79 & 1.01 & 3850 & 0.6&    & 2.05 & 6.89 & $-0.56/  0.08$ & 4.21 & $-0.13 / $\nodata&\tablenotemark{e} & 1.16 & $-0.49 / $\nodata&\tablenotemark{f} & \nodata&\tablenotemark{g} & \nodata$/$\nodata  \\
$J06341077+2421564$ &  9.41 & 1.05 & 3750 & 0.3&    & 2.05 & 6.78 & $-0.67/  0.09$ & 4.27 & $ 0.04 / 0.05$    && 1.23 & $-0.31 / $\nodata&\tablenotemark{f} &  0.30&\tablenotemark{h} & $-0.26 / 0.03$ \\
$J06402752+5944247$ &  9.12 & 1.12 & 3650 & 0.4&    & 1.66 & 7.02 & $-0.43/  0.09$ & 4.12 & $-0.35 / 0.07$    && 1.66 & $-0.12 / 0.10$    &&  0.47&\tablenotemark{f} & $-0.34 / 0.01$ \\
$J06463306+5521444$ & 10.01 & 1.06 & 3750 & 0.9&    & 1.49 & 7.35 & $-0.10/  0.13$ & 4.46 & $-0.34 / 0.00$    && 1.87 & $-0.24 / 0.07$    &&  0.84& & $-0.29 / 0.09$ \\
$J07003934+4730561$ &  8.99 & 1.20 & 3700 & 0.0&(-) & 1.98 & 6.50 & $-0.95/  0.10$ & 4.06 & $ 0.11 / 0.07$    && 1.21 & $-0.05 / 0.02$    &&  0.11& & $-0.17 / 0.02$ \\
$J07022522+2823272$ &  8.63 & 1.21 & 3550 & 0.0&(-) & 1.92 & 6.41 & $-1.04/  0.08$ & 4.13 & $ 0.27 / 0.14$    && 0.55 & $-0.62 / 0.16$    && $-0.08$& & $-0.27 / 0.04$ \\
$J07041061+3911224$ &  7.86 & 1.21 & 3550 & 0.0&(-) & 1.88 & 6.46 & $-0.99/  0.09$ & 4.25 & $ 0.34 / 0.17$    && 1.14 & $-0.08 / 0.13$    && $-0.28$& & $-0.52 / 0.04$ \\
$J07052136+6012353$ &  9.79 & 1.02 & 3800 & 0.8&    & 1.74 & 7.15 & $-0.30/  0.11$ & 4.41 & $-0.19 / 0.01$    && 1.71 & $-0.20 / 0.04$    &&  0.74& & $-0.19 / 0.01$ \\
$J07074384+1319528$ &  9.13 & 1.09 & 3700 & 0.2&    & 2.00 & 6.84 & $-0.61/  0.06$ & 4.43 & $ 0.14 / 0.18$    && 1.16 & $-0.44 / $\nodata&\tablenotemark{f} &  0.43& & $-0.19 / 0.08$ \\
$J07110708+5027276$ &  9.62 & 1.03 & 3800 & 0.8&    & 1.66 & 7.15 & $-0.30/  0.12$ & 4.35 & $-0.25 / 0.01$    && 1.67 & $-0.24 / 0.05$    &&  0.71& & $-0.22 / 0.04$ \\
$J07165539+5409046$ & 10.08 & 1.01 & 3850 & 0.3&    & 1.95 & 6.55 & $-0.90/  0.11$ & 4.14 & $ 0.14 / 0.08$    && 1.15 & $-0.16 / $\nodata&\tablenotemark{d} &  0.17& & $-0.16 / 0.02$ \\
$J07314622+5936181$ &  9.12 & 1.09 & 3700 & 0.0&(-) & 1.96 & 6.44 & $-1.01/  0.08$ & 3.89 & $ 0.00 / 0.03$    && 1.31 & $ 0.07 / 0.12$    &&  0.12& & $-0.10 / 0.01$ \\
$J07393712+3315325$ &  8.85 & 1.14 & 3650 & 0.0&(-) & 1.80 & 6.52 & $-0.93/  0.09$ & 4.17 & $ 0.20 / 0.09$    && 1.02 & $-0.26 / 0.31$    &&  0.10& & $-0.20 / 0.04$ \\
$J08205065+1523392$ &  8.05 & 1.12 & 3700 & 0.0&(-) & 1.60 & 6.51 & $-0.94/  0.06$ & 4.23 & $ 0.27 / 0.06$    && 1.12 & $-0.15 / 0.04$    &&  0.03& & $-0.26 / 0.05$ \\
$J08501690+2545291$ &  9.27 & 1.15 & 3650 & 0.0&(-) & 1.70 & 6.65 & $-0.80/  0.07$ & 3.94 & $-0.16 / 0.08$    && 0.82 & $-0.59 / 0.09$    &&  0.18&   &$ -0.25 / 0.03$ \\
$J08534628+0604280$ &  9.30 & 1.11 & 3700 & 0.0&(-) & 1.64 & 6.53 & $-0.92/  0.07$ & 4.22 & $ 0.24 / 0.00$    && 0.90 & $-0.39 / $\nodata&\tablenotemark{f} & $-0.02$&   & $-0.33 / 0.03$ \\
$J09122898+2632290$ &  9.28 & 1.08 & 3700 & 0.0&    & 1.32 & 6.69 & $-0.76/  0.07$ & 4.40 & $ 0.26 / 0.03$    && 0.96 & $-0.49 / 0.10$    && $-0.14$&   & $-0.61 / 0.04$ \\
$J09164929+5630048$ &  8.98 & 1.11 & 3700 & 0.0&(-) & 1.80 & 6.48 & $-0.97/  0.09$ & 4.12 & $ 0.19 / 0.05$    && 0.91 & $-0.33 / $\nodata&\tablenotemark{f} &  0.04&   & $-0.22 / 0.11$ \\
$J22260469+8218595$ &  9.39 & 1.13 & 3650 & 0.5&    & 1.49 & 7.00 & $-0.45/  0.10$ & 4.03 & $-0.42 / 0.01$    && 1.47 & $-0.29 / $\nodata&\tablenotemark{f} & \nodata&\tablenotemark{g} & \nodata$/$\nodata  \\

\enddata

\tablenotetext{a}{~The effective temperature derived from the
Houdashelt et al. (2000a) color-temperature
relation.}\tablenotetext{b}{~Any entry of the surface gravities
given as ``0.0(-)'' means that our iterative procedure to estimate
the surface gravity (see C07) was converging on a model atmosphere
with $\log{g} < 0$, whereas the model atmosphere grids (Castelli
\& Kurucz 2003) do not go below $\log{g} = 0$, and thus we have
adopted the $\log{g} = 0$ atmosphere in this
case.}\tablenotetext{c}{~With the standard deviation.
}\tablenotetext{d}{~Measurement uncertain due to spectrum defect
on the blue edge of the observed La line.}\tablenotetext{e}{~Only
one \ion{Ti}{+1} line measurable in one order.
}\tablenotetext{f}{~Only one \ion{Y}{+2} line measurable in two
adjacent orders. } \tablenotetext{g}{~Lines unmeasurable due to
the cosmic rays or other defects.} \tablenotetext{h}{~Measurement
uncertain due to unusual shape of the observed La line.}

\end{deluxetable}

\begin{deluxetable}{lccc}
\tabletypesize{\scriptsize} \tablecaption{Sensitivity of
Abundances to Stellar Parameters} \tablewidth{0pt} \tablehead{
\colhead{Star Name} &
 \colhead{$\Delta T_{\rm
eff}=+100$} & \colhead{$\Delta$log {\it g}$=+0.2$} &
\colhead{$\Delta\xi=+0.2$ }
\\
\colhead{Element} &
 \colhead{(K)} & \colhead{(dex)} & \colhead{(${\rm km\,s^{-1}}$) } }

\startdata

$J02340206+8446368$ & & & \\
$\Delta A$(Fe) & $-0.06$ & $+0.07$ & $-0.10$ \\
$\Delta A$(Ti) & $+0.12$ & $+0.03$ & $-0.10$ \\
$\Delta A$(Y) & $-0.04$ & $+0.08$ & $-0.03$ \\
$\Delta A$(La) & $+0.03$ & $+0.08$ & $+0.00$ \\
\\
$J07003934+4730561$ & & & \\
$\Delta A$(Fe) & $-0.07$ & $+0.05$ & $-0.07$ \\
$\Delta A$(Ti) & $+0.15$ & $+0.00$ & $-0.05$ \\
$\Delta A$(Y) & $-0.07$ & $+0.07$ & $-0.03$ \\
$\Delta A$(La) & $+0.01$ & $+0.06$ & $+0.00$ \\
\\
$J07110708+5027276$ & & & \\
$\Delta A$(Fe) & $-0.08$ & $+0.03$ & $-0.10$ \\
$\Delta A$(Ti) & $+0.10$ & $+0.01$ & $-0.07$ \\
$\Delta A$(Y) & $-0.05$ & $+0.06$ & $-0.03$ \\
$\Delta A$(La) & $+0.01$ & $+0.07$ & $+0.00$ \\

\enddata

\end{deluxetable}

\begin{figure}
\plotone{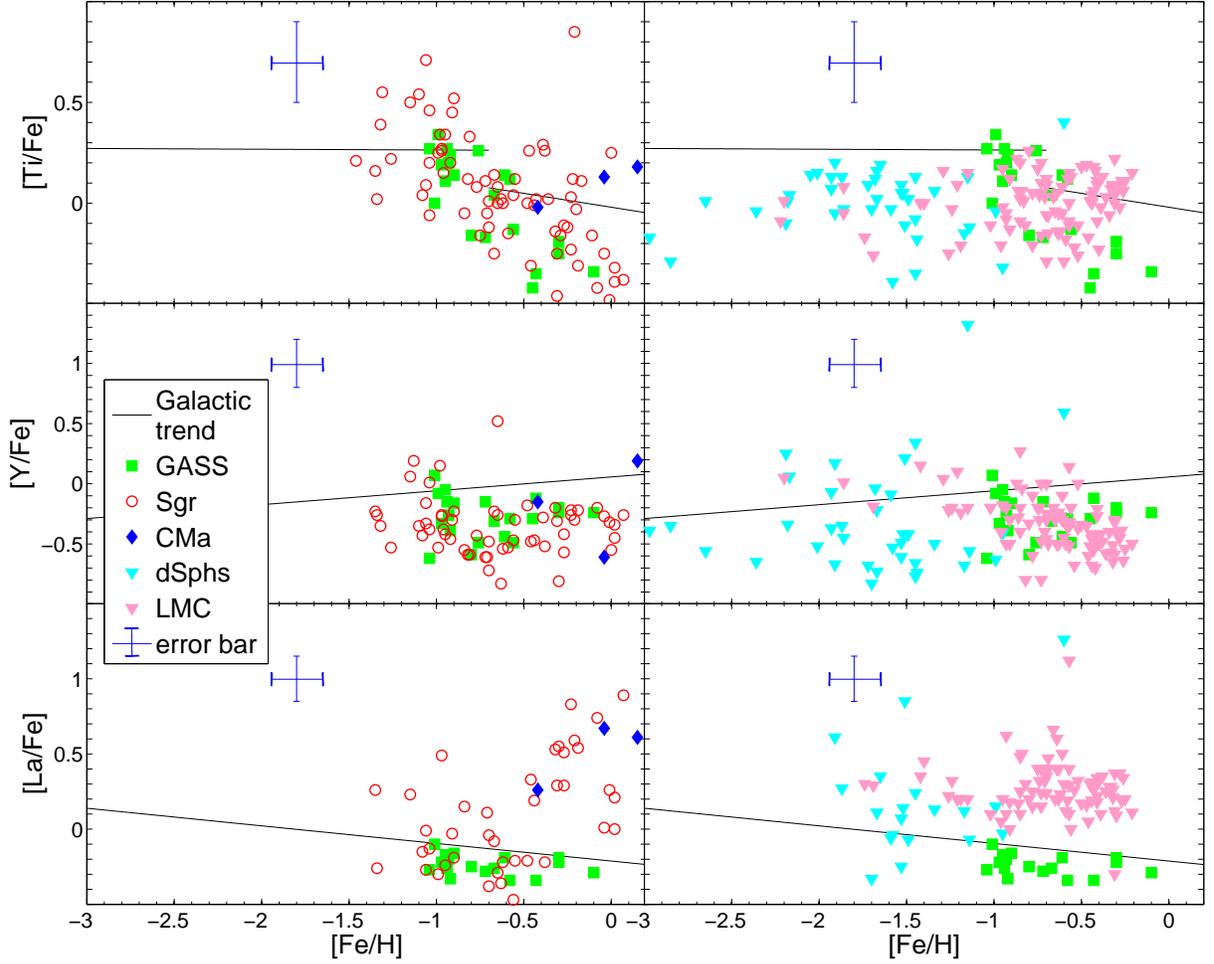} \caption{The distribution as a function of [Fe/H]
of the abundance ratios ({\it top panel}) [Ti/Fe], ({\it middle})
[Y/Fe] and ({\it bottom}) [La/Fe] for GASS stars ({\it green
squares}), comparing with Sgr ({\it red open circles}) and CMa
stars ({\it blue diamonds}) in the left panels, and dSph ({\it
cyan triangles}) and LMC stars ({\it pink triangles}) in the right
panels. The Sgr data come from C10, the CMa data are from Sbordone
et al. (2005), the dSph data are from Shetrone et al.\ (2001;
2003), Sadakane et al.\ (2004) and Geisler et al.\ (2005), the LMC
stars are from Johnson et al.\ (2006), Pomp{\'e}ia et al.\ (2008)
and Mucciarelli et al.\ (2008), and the lines represent the linear
fits to the MW star distribution (see discussion in C10). The MW
data have been taken from Gratton \& Sneden (1994), Fulbright
(2000), Johnson (2002) and Reddy et al.\ (2003). The typical error
bars are shown to the left of each panel in blue.
 }
\end{figure}

\begin{figure}
\plotone{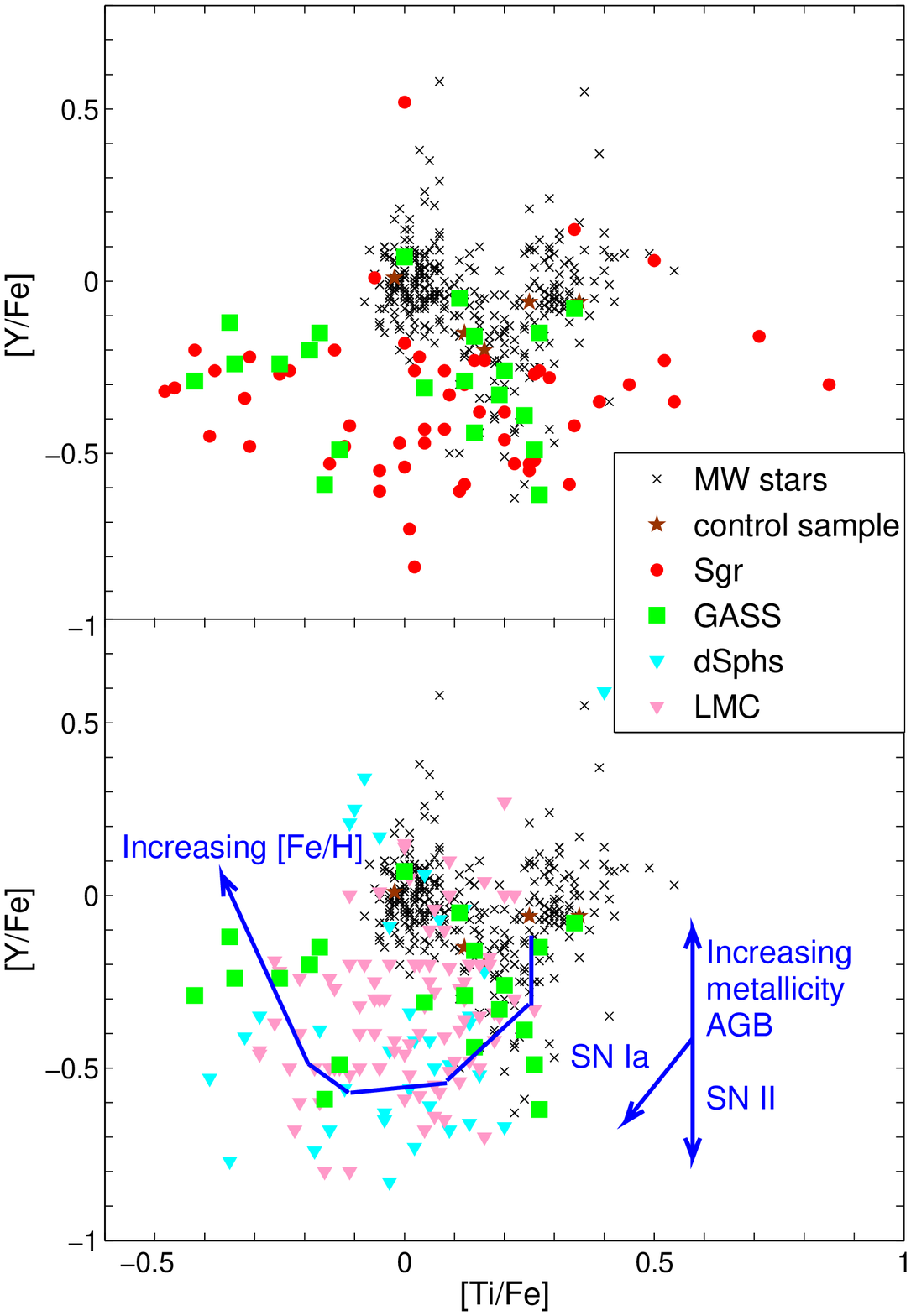} \caption{({\it Top panel}) Distribution of [Y/Fe]
versus [Ti/Fe] for the GASS stars, compared with the Milky Way
(black crosses) and Sgr (red filled circles). The brown star
symbols show the ``control sample'' stars for our survey (C10).
The MW data are from Fulbright (2000), Johnson (2002) and Reddy et
al.\ (2003). The Sgr data are from Sbordone et al. (2007) and C10.
The MW stars lie within the well-defined horseshoe-shaped
distribution near the top. ({\it Bottom panel}) Distribution of
[Y/Fe] versus [Ti/Fe] for the GASS stars, compared with the Milky
Way (black crosses), other dSph (cyan triangles) and LMC stars
(pink triangles). The arrows illustrate schematically which way
various stellar contributions would move the element ratios. }
\end{figure}

\end{document}